\documentclass[
    nature,
    onecolumn,
    10pt,
    nofootinbib,
    superscriptaddress,
    superscriptcite,
    amsfonts,
    amsmath,
    amssymb
]{revtex4-2}

\setcitestyle{super}
\usepackage[utf8]{inputenc}
\usepackage[T1]{fontenc}

\usepackage{graphicx}
\usepackage{dcolumn}
\usepackage{bm}
\usepackage{xcolor}
\usepackage{lipsum}
\usepackage{footmisc}
\usepackage{placeins}
\usepackage{comment}
\usepackage{braket}

\usepackage[
    a4paper,
    left=3cm,
    right=3cm,
    top=2.5cm,
    bottom=3cm
]{geometry}

\usepackage[
    colorlinks=true,
    linkcolor={black},%{blue!80!black},
    citecolor={black},%{blue!80!black},
    urlcolor={black},%{blue!80!black},
]{hyperref}

\renewcommand{\figurename}{Fig.}
\renewcommand{\thefigure}{\arabic{figure}}
\newcommand{\refnum}[1]{Ref.~[\kern-\fontdimen2\font\citenum{#1}]}

\begin{document}

\title{Reversible field-free superconducting diode effect\\controlled by an antiferromagnet}

\author{Filip Krizek}
\affiliation{Institute of Physics, Czech Academy of Sciences, 18200 Prague, Czech Republic}
\affiliation{Faculty of Nuclear Sciences and Physical Engineering, Czech Technical University in Prague, 11519 Prague, Czech Republic}

\author{Kamil Olejník}
\affiliation{Institute of Physics, Czech Academy of Sciences, 18200 Prague, Czech Republic}

\author{Tobias Edwinson}
\affiliation{Niels Bohr Institute, University of Copenhagen, 2100 Copenhagen, Denmark}

\author{August Jacobi}
\affiliation{Niels Bohr Institute, University of Copenhagen, 2100 Copenhagen, Denmark}

\author{Athira Suresh}
\affiliation{Niels Bohr Institute, University of Copenhagen, 2100 Copenhagen, Denmark}

\author{Andrej Farkaš}
\affiliation{Institute of Physics, Czech Academy of Sciences, 18200 Prague, Czech Republic}
\affiliation{Faculty of Mathematics and Physics, Charles University, 121 16 Prague, Czech Republic}

\author{Jan Kraus}
\affiliation{Institute of Physics, Czech Academy of Sciences, 18200 Prague, Czech Republic}
\affiliation{Faculty of Mathematics and Physics, Charles University, 121 16 Prague, Czech Republic}

\author{Vít Novák}
\affiliation{Institute of Physics, Czech Academy of Sciences, 18200 Prague, Czech Republic}

\author{Christoph Müller}
\affiliation{Institute of Physics, Czech Academy of Sciences, 18200 Prague, Czech Republic}
\affiliation{Faculty of Mathematics and Physics, Charles University, 121 16 Prague, Czech Republic}

\author{Vojtěch Pařízek}
\affiliation{Institute of Physics, Czech Academy of Sciences, 18200 Prague, Czech Republic}
\affiliation{Faculty of Mathematics and Physics, Charles University, 121 16 Prague, Czech Republic}

\author{Niclas Heinsdorf}
\affiliation{Department of Physics and Institute for Quantum Information and Matter, California Institute of Technology, Pasadena CA 91125, USA}

\author{Peter Wadley}
\affiliation{School of Physics and Astronomy, University of Nottingham, Nottingham NG7 2RD, United Kingdom}

\author{Oliver Amin}
\affiliation{School of Physics and Astronomy, University of Nottingham, Nottingham NG7 2RD, United Kingdom}

\author{Kevin Edmonds}
\affiliation{School of Physics and Astronomy, University of Nottingham, Nottingham NG7 2RD, United Kingdom}

\author{Tomas Jungwirth}
\affiliation{Institute of Physics, Czech Academy of Sciences, 18200 Prague, Czech Republic}
\affiliation{School of Physics and Astronomy, University of Nottingham, Nottingham NG7 2RD, United Kingdom}
\affiliation{Center for Science and Innovation in Spintronics, Tohoku University, Sendai, Japan}

\author{Libor Šmejkal}
\affiliation{Institute of Physics, Czech Academy of Sciences, 18200 Prague, Czech Republic}
\affiliation{Max Planck Institute for the Physics of Complex Systems, 01187 Dresden, Germany}

\author{Anna Birk Hellenes}
\affiliation{Institute of Physics, Czech Academy of Sciences, 18200 Prague, Czech Republic}

\author{Sumit Ghosh}
\affiliation{Institute of Physics, Czech Academy of Sciences, 18200 Prague, Czech Republic}

\author{Michal Mazur}
\affiliation{Faculty of Science, Charles University, 12843 Prague, Czech Republic}

\author{Dominik Kriegner}
\affiliation{Institute of Physics, Czech Academy of Sciences, 18200 Prague, Czech Republic}

\author{Lucas Casparis}
\affiliation{Niels Bohr Institute, University of Copenhagen, 2100 Copenhagen, Denmark}

\author{Saulius Vaitiek\.{e}nas}
\affiliation{Niels Bohr Institute, University of Copenhagen, 2100 Copenhagen, Denmark}

\date{\today}
\maketitle

\begin{center}
\begin{minipage}{0.8\textwidth}
\normalsize
The semiconductor diode, which allows current to flow preferentially in one direction, is a fundamental building block of numerous modern electronic circuits. 
Its superconducting analogue---the superconducting diode effect~\cite{Ando2020}---enables directional dissipationless current flow and may provide similar functionality in future superconducting quantum circuits~\cite{Nadeem2023}.
Realization of such a nonreciprocal supercurrent requires broken time-reversal symmetry~\cite{Nagaosa2023}.
At zero applied field, this has been typically associated either with intrinsic unconventional superconductivity~\cite{Lin2022,Le2024,Banerjee2024a,Chakraborty2025} or extrinsic spin-split electronic states induced by magnetic proximity~\cite{Narita2022,Jeon2022,Banerjee2024a,Sachin2026}.
Here we demonstrate a field-free superconducting diode effect in conventional superconducting Al proximitized by collinear antiferromagnetic CuMnAs, whose electronic structure breaks time-reversal symmetry without generating spin splitting.
By tuning the proximity effect through an insulating AlAs interlayer and correlating the reversal of the diode polarity with the reversal of the remanent Néel state, we establish that the antiferromagnet controls the superconducting diode effect.
Our results show that neither spin-split bands nor net magnetization is required for a magnetically controlled field-free superconducting diode effect, extending superconducting nonreciprocity to a broader class of collinear compensated magnets.
\end{minipage}
\end{center}

%%%%%%%%%%%%%%%%%%%%% FIG. 1 %%%%%%%%%%%%%%%%%%%%%%%
\begin{figure}[t!]
    \centering
    \includegraphics[width = 0.8\linewidth]{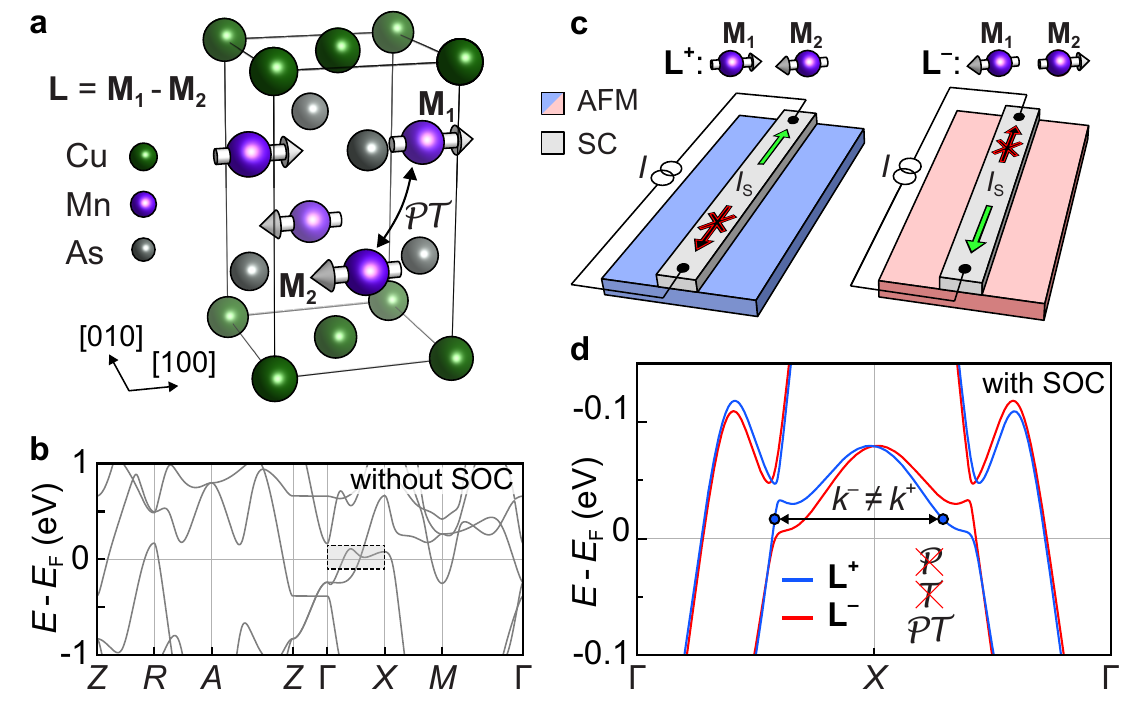}
    \caption{ 
   \textbf{Symmetry of CuMnAs and the superconducting diode effect.} 
   \textbf{a,} Schematic illustration of the CuMnAs unit cell. Arrows indicate the Mn magnetic moments of the two sub-lattices $\mathbf M_1$ and $\mathbf M_2$, with the Néel vector defined as $\mathbf L = \mathbf  M_1 - \mathbf M_2$.
   The states $L^+$ and $L^-$ correspond to opposite Néel-vector orientations along the [100] axis.
   \textbf{b,}~Non-relativistic density functional theory band structure of CuMnAs without spin-orbit coupling (SOC). The gray box marks the Dirac crossing highlighted in \textbf{d}.
   \textbf{c,}~Schematic of the superconducting diode effect in a hybrid $\cal{PT}$-symmetric antiferromagnet/superconductor bar, illustrating reversal of the diode polarity upon reversal of the Néel vector ($\mathbf L^+ \leftrightarrow \mathbf L^-$).
   \textbf{d},~Relativistic band structure near the Fermi level with SOC. The two-fold spin degeneracy remains protected by $\cal{PT}$ symmetry. The blue and red curves correspond to opposite Néel-vector orientations, $\mathbf L^+$ and $\mathbf L^-$, rather than opposite spins.
   The individually broken $\cal{P}$ and $\cal{T}$ symmetries are reflected in an asymmetric momentum splitting of the Dirac points at time-reversed momenta. Reversing the Néel vector reverses the sense of this band-structure asymmetry.
    }
    \label{f1}
\end{figure}
%%%%%%%%%%%%%%%%%%%%%%%%%%%%%%%%%%%%%%%%%%%%%%%%%%%%

The time-reversal ($\cal{T}$) symmetry operation is of fundamental importance in physics. Already in 1932, Wigner provided a symmetry interpretation~\cite{Wigner1932} of the Kramers theorem~\cite{Kramers1930} linking the spin degeneracy of electronic states in atoms to $\cal{T}$-symmetry. 
This naturally led to the common association between spontaneously broken $\cal{T}$ symmetry in the electronic band structure of magnets and lifting of their spin degeneracy.  
This perspective has motivated researchers to consider altermagnets as promising candidates for realizing the superconducting diode effect (SDE) without applied magnetic field and magnetization~\cite{Banerjee2024a}, because they combine $\cal{T}$-symmetry-breaking spin-split electronic structure with vanishing net magnetization of their compensated collinear magnetic order~\cite{Smejkal2021a,Smejkal2022a}. 

Experimentally, a field-free SDE was recently reported in the compensated magnet Mn$_3$Pt, whose more complex multi-sublattice magnetic order gives rise to a $\cal{T}$-symmetry-breaking spin-split electronic structure that also features a non-collinear momentum-dependent spin texture~\cite{Sachin2026}.
However, reversible control of the magnetic order together with the diode polarity remained elusive in that system, leaving their microscopic connection open and limiting direct application in quantum circuits.

In this work, we use semimetallic CuMnAs featuring a common two-sublattice collinear antiferromagnetic order, characterized by the Néel vector $\mathbf L=\mathbf M_1-\mathbf M_2$, as shown in Figs.~\ref{f1}a,b. 
This compensated magnet allows us to demonstrate reversible control of the field-free and magnetization-free superconducting diode polarity by manipulating the remanent N\'eel state, as schematically illustrated in Fig.~\ref{f1}c. 

Remarkably, antiferromagnetic CuMnAs provides the required broken $\cal{T}$ symmetry of the electronic states while retaining fully spin-degenerate band structure, as evidenced by first-principles calculations (Fig.~\ref{f1}d). 
In fact, CuMnAs is the magnet in which this departure from the usual association between broken $\cal{T}$ symmetry and spin splitting was identified~\cite{Tang2016, Smejkal2017c}. 
Here, instead of Kramers degeneracy associated with $\cal{T}$-symmetry, the spin degeneracy of the band structure is protected by the combined $\cal{PT}$ symmetry of the antiferromagnetic crystal (Fig.~\ref{f1}a), where $\cal{P}$ denotes space inversion. 
The $\cal{PT}$ symmetry, and the resulting spin degeneracy, also protect precisely zero net magnetization. 
Simultaneously, both $\cal{P}$ and $\cal{T}$ symmetries are individually broken in the electronic band structure by the collinear antiferromagnetic order of CuMnAs (Figs.~\ref{f1}a,d).  

The broken $\cal{T}$ and $\cal{P}$ symmetries are particularly apparent in the relativistic band structure near the Fermi level, containing Dirac points at time-reversed momenta (Fig.~\ref{f1}d). 
Without spin-orbit coupling (SOC), $\cal{PT}$ symmetry protects the four-fold degeneracy of the Dirac crossing (highlighted by the gray box in Fig.~\ref{f1}b).
With SOC, the two-fold spin degeneracy of the band structure remains protected by the $\cal{PT}$ symmetry, while the individually broken $\cal{T}$ and $\cal{P}$ symmetries are reflected in the asymmetric splitting of the Dirac points. 
The sense of this band-structure asymmetry reverses when the antiferromagnetic N\'eel vector is reversed (Fig.~\ref{f1}d).

%%%%%%%%%%%%%%%%%%%%% FIG. 2 %%%%%%%%%%%%%%%%%%%%%%%
\begin{figure}
    \centering
    \includegraphics[width = 0.8\linewidth]{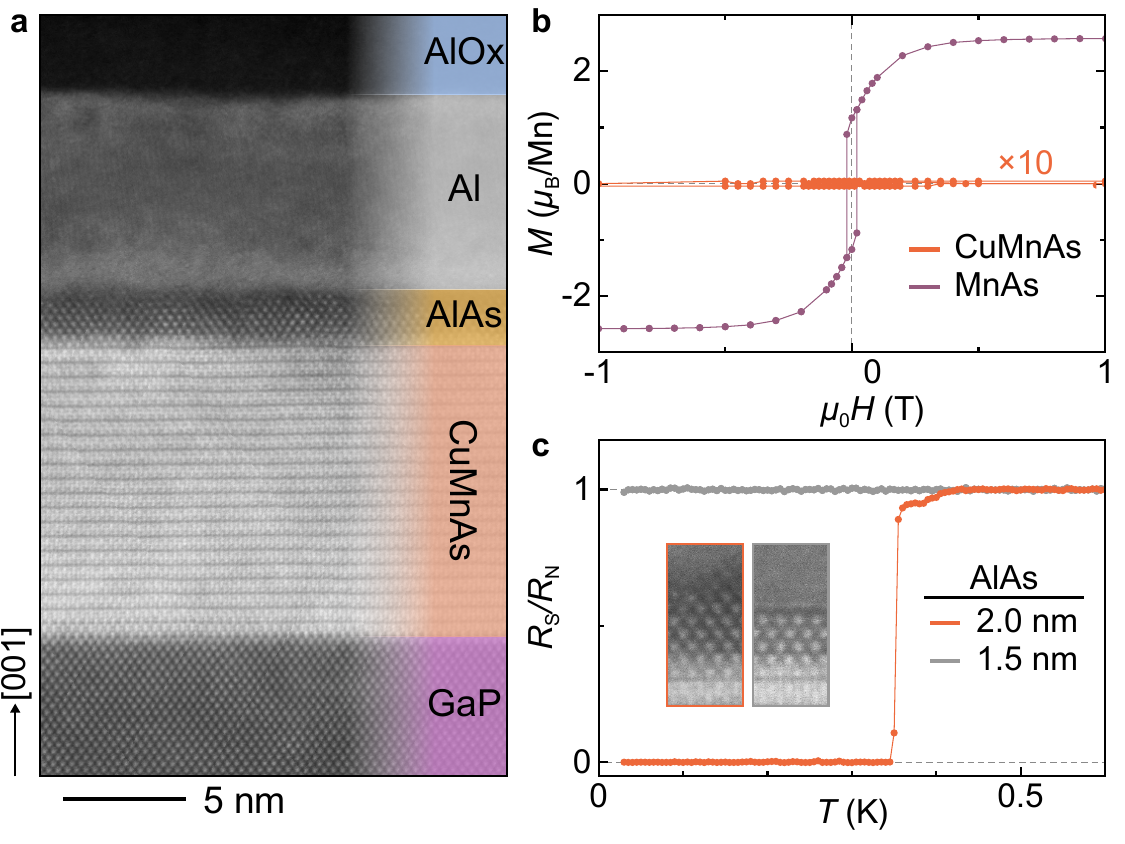}
    \caption{ 
    \textbf{CuMnAs/AlAs/Al heterostructure and proximity tuning.} 
    \textbf{a,} High-angle annular dark-field scanning transmission electron microscopy (STEM) image of the epitaxial CuMnAs/AlAs/Al heterostructure on GaP. The scale bar is 5 nm.
    \textbf{b,} Magnetization per Mn atom, $M$, as a function of applied magnetic field, $H$, for the CuMnAs heterostructure and a reference ferromagnetic MnAs sample. The CuMnAs signal is multiplied by 10.
    \textbf{c,} Normalized sheet resistance, $R_{\rm S}/R_{\rm N}$, as a function of temperature, $T$, for heterostructures with 2.0 and 1.5~nm AlAs barriers. 
    Superconductivity is recovered for the 2.0~nm barrier, with $T_{\rm C}\approx350$~mK, whereas no transition is observed for the 1.5 nm barrier down to 20~mK. Insets: enlarged STEM images of the AlAs interfaces for the two barrier thicknesses.
    }
    \label{f2}
\end{figure}
%%%%%%%%%%%%%%%%%%%%%%%%%%%%%%%%%%%%%%%%%%%%%%%%%%%%

Over the past decade, semimetallic CuMnAs has been a workhorse material in antiferromagnetic spintronics~\cite{Jungwirth2016}, enabling all-electrical antiferromagnetic memory~\cite{Wadley2016} or terahertz switching~\cite{Olejnik2018}, among other advances. 
N\'eel-vector reversal has been observed directly by microscopic domain imaging~\cite{Janda2020a, Amin2023} and detected electrically through nonlinear magnetotransport~\cite{godinho2018electrically}. 
The mature molecular-beam-epitaxy (MBE) growth of high-quality CuMnAs thin films benefits from the prevailing strong non-metallic nature of bonding,  enabling atomically sharp non-intermixed interfaces in epitaxial heterostructures~\cite{Wadley2013,Krizek2020}. 
In our experiments, the heterostructure grown by MBE on (001) GaP substrate consists of a 12.4~nm CuMnAs single-crystalline epilayer, a 2~nm self-assembled AlAs barrier, and a 9.2~nm Al epilayer. 
The layer structure and epitaxial quality of the entire heterostructure are shown in the scanning transmission electron microscopy (STEM) image in Fig.~\ref{f2}a (for additional structural characterization, see Extended Data Figs.~\ref{edf1} and \ref{edf2}).
The vanishingly small net magnetization of our CuMnAs film is confirmed by the superconducting quantum interference device (SQUID) measurements (Fig.~\ref{f2}b), which place an upper bound of ($3\times10^{-3}~\mu_B$/Mn), more than an order of magnitude below that expected from a uniformly uncompensated interfacial Mn plane~\cite{Wadley2017}. 
The N\'eel temperature of antiferromagnetic CuMnAs is 480~K (see \refnum{wadley2015antiferromagnetic}), far above the 1.2~K nominal superconducting transition temperature of Al.

%%%%%%%%%%%%%%%%%%%%% FIG. 3 %%%%%%%%%%%%%%%%%%%%%%%
\begin{figure}[bt!]
    \centering
    \includegraphics[width = 0.8\linewidth]{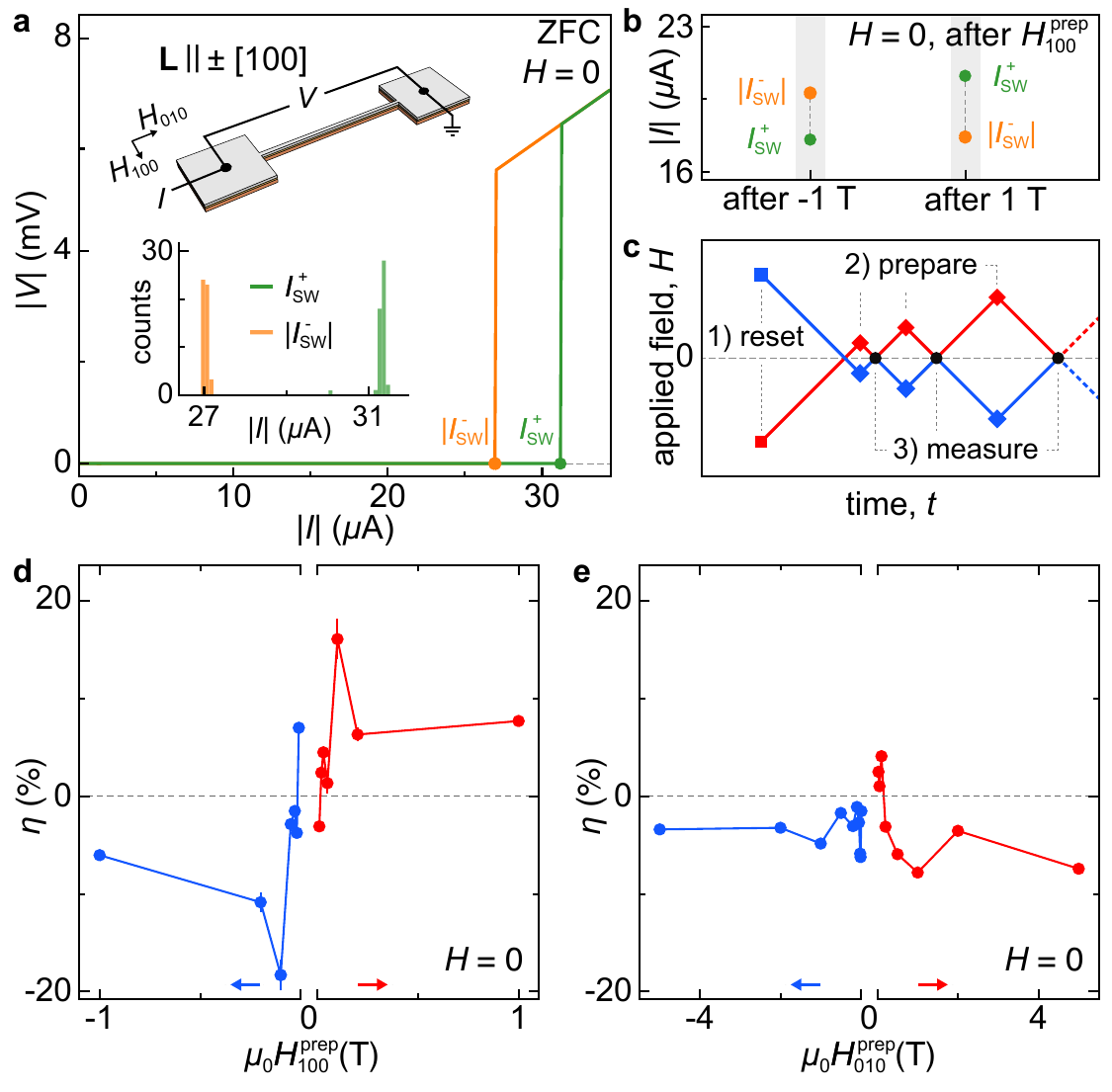}
    \caption{ 
    \textbf{Remanent control of the field-free superconducting diode effect.}
    \textbf{a,}~Absolute voltage, $\vert V\vert$, as a function of current magnitude, $\vert I\vert$, measured after zero-field cooling (at $H=0$). 
    Top inset: schematic of the device geometry overlaid with four-terminal measurement configuration.
    Bottom inset: switching-current, $I_{\mathrm{SW}}$, distributions for positive and negative current sweeps. 
    \textbf{b,}~Switching currents measured at $H=0$ after preparation fields $H^{\mathrm{prep}}_{100}=\pm1$~T. Reversing the preparation field interchanges $I_{\mathrm{SW}}^{+}$ and $\vert I_{\mathrm{SW}}^{-}\vert$.
    \textbf{c,}~Magnetic-field preparation protocol. The device is reset, exposed to a finite preparation field, and measured after returning to $H=0$. Red and blue indicate opposite preparation-field polarities.
    \textbf{d, e,}~Zero-field superconducting diode efficiency, $\eta$, after field preparation along \textbf{d} [100] and \textbf{e} [010] crystallographic directions. Preparation along [100] reverses the diode polarity, whereas preparation along [010] produces a weaker response and leaves $\eta$ predominantly of the same sign.
    }
    \label{f3}
\end{figure}
%%%%%%%%%%%%%%%%%%%%%%%%%%%%%%%%%%%%%%%%%%%%%%%%%%%%

Without increasing the Al thickness, we tune the strength of the proximity coupling between antiferromagnetic CuMnAs and superconducting Al by introducing a thin insulating AlAs barrier between the two layers. 
For 1.5~nm AlAs, superconductivity in Al is suppressed down to the lowest measured temperature of 20~mK (Fig.~\ref{f2}c).
Increasing the AlAs thickness to 2~nm~(Fig.~\ref{f2}a) reduces the coupling between CuMnAs and Al sufficiently to recover superconductivity. 
The resulting critical temperature of 350~mK nevertheless remains substantially below the nominal 1.2~K transition temperature of Al (Fig.~\ref{f2}c), indicating that a sizable proximity coupling is retained in the heterostructure. 
This behavior is consistent with calculations showing a crossover from a strongly suppressed to a sizable superconducting state as the spacer thickness is increased (Extended Data Fig.~\ref{edf3}).

We next demonstrate how the superconducting transport inherits the broken time-reversal symmetry of the antiferromagnetic CuMnAs layer. 
To this end, we measure the current–voltage characteristics of a $5~\mu$m-wide and $180~\mu$m-long CuMnAs/Al bar in a four-terminal configuration, as schematically illustrated in Fig.~\ref{f3}a.
After zero-field cooling (ZFC) from above the superconducting transition temperature of Al, the device switches from the superconducting to the resistive state at different currents for opposite current polarities, $I_{\mathrm{SW}}^{+}\neq \vert I_{\mathrm{SW}}^{-}\vert$, confirming a field-free superconducting diode effect. 
We quantify the nonreciprocity by the diode efficiency,
$\eta=\left(I_{\mathrm{SW}}^{+}-\vert I_{\mathrm{SW}}^{-}\vert\right)/\left(I_{\mathrm{SW}}^{+}+\vert I_{\mathrm{SW}}^{-}\vert\right)$, which reaches approximately $7\%$ in the ZFC state. 
Repeated current sweeps yield well-separated switching-current distributions for the two current directions (inset of Fig.~\ref{f3}a), showing that the observed asymmetry well exceeds the stochastic spread of the switching process. 

To demonstrate that the zero-field diode state can be controlled magnetically, we apply a finite preparation field, $H^{\mathrm{prep}}$, and return the field to zero ($H=0$) before measuring the switching currents. 
For opposite preparation field values along [100], $H^{\mathrm{prep}}_{100}=\pm1$~T, we observe that the ordering of the two switching currents is reversed: $I_{\mathrm{SW}}^{+}<\vert I_{\mathrm{SW}}^{-}\vert$ after a negative $H^{\mathrm{prep}}_{100}$ excursion, whereas $I_{\mathrm{SW}}^{+}>\vert I_{\mathrm{SW}}^{-}\vert$ after a positive $H^{\mathrm{prep}}_{100}$ excursion (Fig.~\ref{f3}b).
We note that the two prepared states retain a comparable mean switching-current scale, showing that the preparation field predominantly reverses the nonreciprocal component rather than simply strengthening or suppressing superconductivity.

We map this remanent response using a field-excursion protocol (Fig.~\ref{f3}c): after resetting the system at high field, we apply the preparation field, $H^{\rm prep}$, of opposite sign and progressively increasing its magnitude, reading out the diode efficiency at $H=0$ after each excursion. 
Applying the resetting and preparation field along [100], $\eta$ evolves continuously with the preparation history and reverses sign between opposite field polarities (Fig.~\ref{f3}d), reaching values approaching $\pm20\%$. 
In contrast, applying the resetting and preparation fields along the orthogonal $[010]$ direction produces a substantially weaker response which shows no systematic reversal of the sign of the diode effect (Fig.~\ref{f3}e). 
The field-free SDE is therefore not only remanent but also strongly sensitive to the sign and crystal direction along which the field is applied before measurement.

The measurements in Fig.~\ref{f3} were performed on a bar oriented along the [010] crystallographic direction.
An orthogonal bar patterned along [100] displays the same qualitative dependence on the preparation-field axes (Extended Data Fig.~\ref{edf4}). 
This suggests that the anisotropy is not determined by the relative orientation of the applied field and current, but is instead tied to the underlying crystallographic axes of CuMnAs.
If so, the same crystallographic anisotropy should also be apparent directly in the antiferromagnetic response.
We therefore perform nonlinear magnetotransport measurements at 5~K using a reference CuMnAs bar without a superconducting Al layer. 
Because $\cal{P}$ and $\cal{T}$ are individually broken in CuMnAs, the nonlinear resistive response is sensitive to the orientation of the antiferromagnetic order.
In particular, unlike conventional linear magnetoresistance, the even-order nonlinear response changes sign under Néel-vector reversal and has previously been used to electrically distinguish oppositely oriented Néel states~\cite{godinho2018electrically}. This provides an independent probe of how the CuMnAs magnetic configuration responds to, and retains memory of, the field-preparation protocol, without invoking superconducting transport.

\begin{figure}[bt!]
    \centering
    \includegraphics[width = 0.8\linewidth]{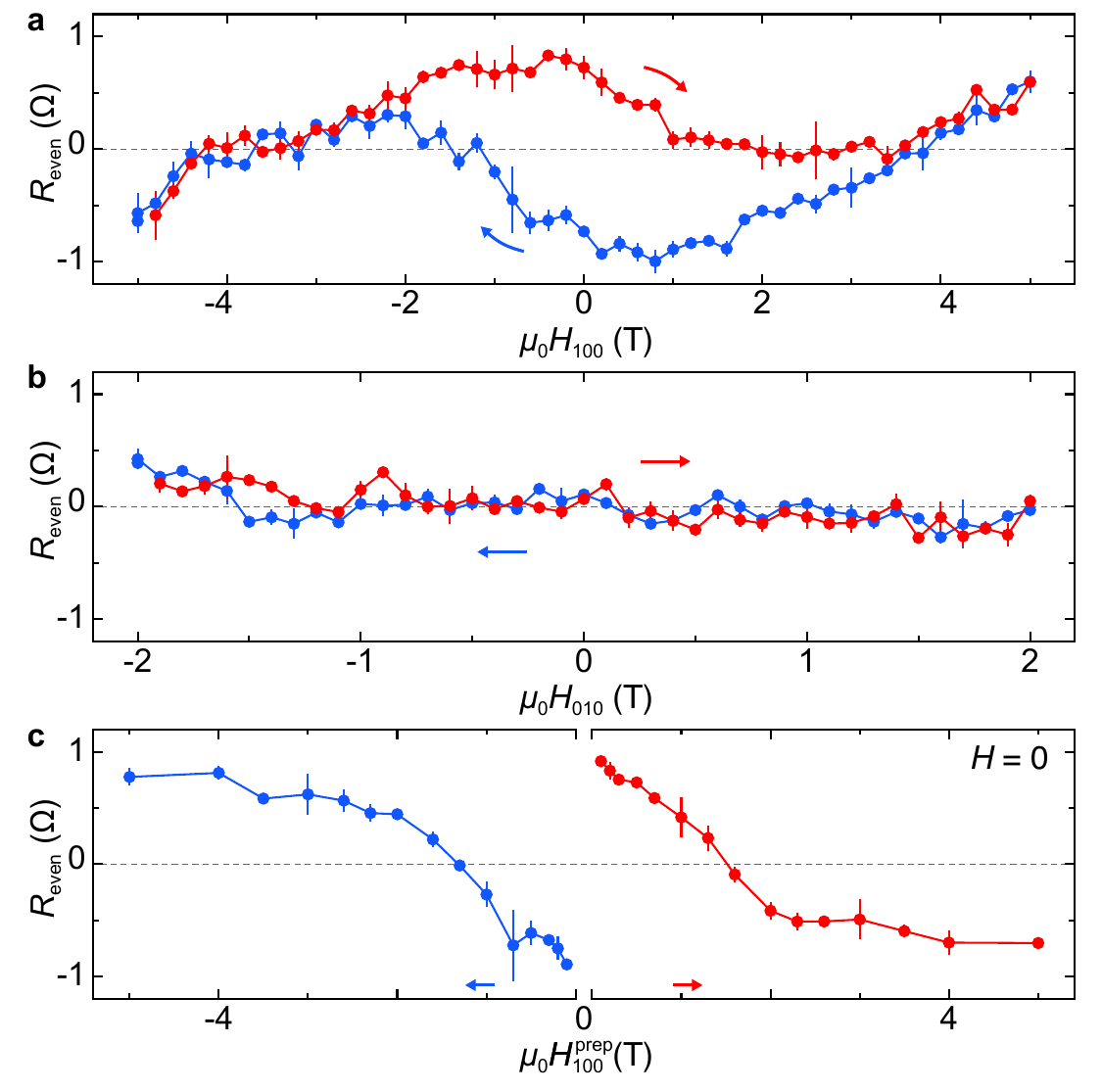}
    \caption{ 
    \textbf{Nonlinear transport probe of Néel-vector manipulation.}
    \textbf{a,} Even-in-current nonlinear resistance, $R_{\rm even}=[V(+I)+V(-I)]/(2I)$, measured as a function of magnetic field applied along [100]. The response is pronounced and hysteretic.
    \textbf{b,} Corresponding measurement for magnetic field applied along [010], showing a substantially weaker response.
    \textbf{c,} Remanent $R_{\rm even}$ measured at $H=0$ after magnetic-field preparation along [100], following the protocol of Fig.~\ref{f3}c. Opposite preparation-field polarities produce opposite zero-field nonlinear responses.
    The data were taken at 5~K.
    }
    \label{f4}
\end{figure}

We characterize the even-in-current nonlinear resistance as $R_{\mathrm{even}}=[V(+I)+V(-I)]/(2I)$.
Sweeping the magnetic field along [100] produces a pronounced and hysteretic nonlinear response, whereas the field response along [010] is substantially weaker and remains close to zero (Fig.~\ref{f4}a,b). 
This anisotropy mirrors that of the remanent SDE and is independently supported by X-ray magnetic linear dichroism photoemission electron microscopy (XMLD-PEEM) measurements (Extended Data Fig.~\ref{edf5}), which reveal a uniaxial easy axis along [100] with only a weak biaxial contribution from the substrate surface anisotropy~\cite{wang2020spin}.

%%%%%%%%%%%%%%%%%%%%% FIG. 4 %%%%%%%%%%%%%%%%%%%%%%%
%%%%%%%%%%%%%%%%%%%%%%%%%%%%%%%%%%%%%%%%%%%%%%%%%%%%

To test whether this correspondence extends to the remanent magnetic state, we apply the same field-preparation protocol used for the diode measurements~(Fig.~\ref{f3}c) and read out $R_{\mathrm{even}}$ only after returning to $H=0$ (Fig.~\ref{f4}c).
For $H_{100}^{\mathrm{prep}}$ ranging from $-5$ to 5~T, $R_{\mathrm{even}}$ shows pronounced hysteresis, with opposite preparation-field polarities producing opposite zero-field nonlinear responses. 
Since the nonlinear response is sensitive to Néel-vector reversal, these two branches indicate opposite imbalances between oppositely oriented Néel-domain populations retained after the field is removed. 
Their polarity dependence closely mirrors the reversal of the remanent diode efficiency in Fig.~\ref{f3}d, providing an independent magnetic signature of the field-written states underlying the superconducting diode response.

Taken together, the shared crystallographic anisotropy and matching remanent field-history dependence establish a direct correspondence between the antiferromagnetic state of CuMnAs and the field-free diode response.
Importantly, the memory effect of the non-linear magnetoresistance we observe in a reference structure without a superconducting layer shows that the anisotropic remanence itself originates in CuMnAs rather than from spurious flux trapping or vortex dynamics, both known to generate SDE~\cite{Hou2023, Jiang2025}. 
Together with the vanishingly small residual magnetic moment measured using SQUID (Fig.~\ref{f2}b), this also disfavors a stray-field mechanism associated with uncompensated magnetization~\cite{Li2026}. 
Neither mechanism is naturally expected to track the Néel-vector anisotropy observed here. 
We therefore conclude that the polarity of the field-free SDE is controlled by the remanent antiferromagnetic state of CuMnAs.

This control mechanism differs fundamentally from previously proposed realizations of a field-free SDE. 
In altermagnet~\cite{Banerjee2024a,Smejkal2022a}, non-collinear compensated-magnet~\cite{Sachin2026}, and proximitized-ferromagnet hosts~\cite{Narita2022, Jeon2022}, the broken $\cal T$ symmetry is accompanied by spin-split electronic band structure. 
In CuMnAs, by contrast, $\cal PT$ symmetry protects spin degeneracy of the band structure even though $\cal P$ and $\cal T$ are individually broken (Fig.~\ref{f1}). 
Our results therefore show that spin-split bands are not a prerequisite for a magnetically controlled field-free SDE. 
Instead, a collinear compensated antiferromagnet with a spin-degenerate electronic structure can imprint reversible superconducting nonreciprocity onto a conventional s-wave superconductor.

Because the diode polarity is encoded in a remanent antiferromagnetic state rather than requiring a sustained external field or net magnetization, this platform provides a route towards non-volatile superconducting nonreciprocity with minimal stray fields. 
Collinear $\cal{PT}$-symmetric antiferromagnets are comparatively abundant, and their Néel order can already be written electrically at terahertz speeds~\cite{Wadley2016,Olejnik2018} and imaged non-destructively~\cite{Janda2020a,godinho2018electrically}.
Combining these established antiferromagnetic-spintronic capabilities with conventional superconductors could therefore enable reconfigurable superconducting diodes and phase-biased Josephson elements whose state is stored in the antiferromagnetic state.\\

\textbf{METHODS}

\textbf{Material Growth.}
The CuMnAs/AlAs/Al heterostructures shown in Fig.~\ref{f2}a were grown by molecular beam epitaxy (MBE) on (001) GaP substrates. 
An epi-ready 2" GaP wafer was desorbed at 640$^{\circ}\text{C}$ under P overpressure (temperatures were measured by band-edge spectroscopy\cite{novak2007substrate}). 
The 80 nm thick GaP buffer layer was grown at 550$^{\circ}\text{C}$ with 1.3:1 P:Ga ratio at 0.2 \AA $/$s. 
After cooling down, the nominally 13~nm thick CuMnAs layer was grown at 220$^{\circ}\text{C}$ with all three individual fluxes tuned to achieve 1:1:1 stoichiometry at 0.05 \AA $/$s. 
We note that, unlike \refnum{Krizek2020}, the film is free of extended crystallographic defects, due to precise tuning of the As$_4$ flux (see Extended Data Fig.~\ref{edf2}). 
The nominally 12 nm thick Al film was grown below 30$^{\circ}\text{C}$ at 0.3~\AA $/$s. 
The substrate was passively cooled down by the MBE cryopanel to below -17$^{\circ}\text{C}$ and warmed up close to 30$^{\circ}\text{C}$ due to cell radiation during growth. 
The 2 nm thick AlAs interlayer is self-assembled during Al growth by alloying Al with a thin As layer condensed from As$_4$ background pressure during the cooldown step. 
For the reference sample with a 1.5~nm AlAs barrier, the self-assembled barrier thickness was reduced by removing the sample from the growth chamber into ultra-high vacuum immediately after CuMnAs growth. 
It was reintroduced after the chamber background pressure reached \mbox{$10^{-9}$ Pa}, and the Al film was then grown as above. 

\textbf{Device fabrication.} 
The heterostructure with a 2.0~nm AlAs barrier was patterned into strips by UV lithography using AZ 1505 photoresist. 
The Al top layer was first removed by wet etching in a TMAH-based developer (AR 300-475) for 45~s at room temperature. 
Using the same resist mask, the exposed AlAs and CuMnAs layers were then removed by Kaufmann argon-ion milling in an AJA Orion system for 5 min at a beam voltage of 300~V, a beam current of 23~mA, a pressure of 1~mTorr, and an incident angle of 0$^\circ$. 
The resist was subsequently stripped in dioxolane.

\textbf{DFT calculations.} 
The density functional theory (DFT) calculations were performed with the plane-wave Vienna ab initio Simulation Package (VASP)~\cite{Kresse1996a} using the projector-augmented-wave method~\cite{Blochl1994,Kresse1999} and the Perdew-Burke-Ernzerhof (PBE) exchange-correlation functional~\cite{Perdew1996}. 
A Hubbard $U$ correction was applied to the Mn $d$ orbitals in the Dudarev scheme \cite{Dudarev1998}, $U-J=1.7\,\mathrm{eV}$, comparable to \refnum{Veis2018}.
Calculations were performed both with and without spin-orbit coupling. 
An energy cut-off of $400\, \mathrm{eV}$ was used, and all calculations were converged within $10^{-7}\, \mathrm{eV}$.
The lattice constants we used were \mbox{$a=b=3.80$ \AA}  and $c=6.32$ \AA, and a $16\times 16\times 10$ $k$-grid was used for the self-consistent calculation.

\textbf{Superconducting-instability calculation.} 
To assess the role of the spacer thickness on the superconductivity in our system, we model the Al/AlAs/CuMnAs heterostructure by an effective tight-binding stack comprising 41 superconductor layers, $N_{\rm S}=1,\ldots,8$ semiconductor spacer layers and a single effective antiferromagnetic layer, with open boundaries along $z$ and a common periodic square cell in plane~\cite{Heinsdorf2026, Liudeng2026}.
All energies are given in eV relative to $E_F=0$, with dimensionless $\mathbf{k}\in[-\pi,\pi]^2$. The superconductor has one orbital per layer with dispersion $\xi_{\mathrm{Al}}=-2(\cos k_x+\cos k_y)+2$.
The semiconductor has uncoupled valence and conduction orbitals with layer energies $\varepsilon_v=-1.08-0.8(2-\cos k_x-\cos k_y)-2t_{s,z}$ and $\varepsilon_c=1.08+(1+\cos k_x\cos k_y)+2t_{s,z}$, where $t_{s,z}=5$.
These parameters correspond to a semiconductor band gap of $2.16$ eV.
The antiferromagnet is represented by a minimal two-sublattice layer with opposite exchange fields, intended to capture the magnetic pair-breaking environment.
Its sublattices $A,B$ have onsite energies $(-4+0.5\sigma)$ and $(-4-0.5\sigma)$, respectively, for spin $\sigma=\pm1$, and intersublattice Bloch hopping $g(\mathbf{k})=-(1+e^{-ik_x})(1+e^{-ik_y})$ and its Hermitian conjugate.
Nearest-layer hopping amplitudes are $-3$ in the superconductor and $+5$ and $-5$ for the semiconductor valence and conduction orbitals, respectively.
Spin-conserving interface hopping couples the superconductor equally to both semiconductor orbitals with amplitude $-2.6/\sqrt{2}$, and couples $v\leftrightarrow A$ and $c\leftrightarrow B$ with amplitude $-2.6$.
Only $N_{\rm S}$ is varied to represent the spacer thickness; taking one (001) AlAs monolayer as approximately $0.283~\mathrm{nm}$, the experimentally studied $1.5$ and $2.0~\mathrm{nm}$ barriers correspond to approximately 5 and 7 monolayers, respectively.

At $T=1$~K, we use an instantaneous on-site singlet attraction $U=1.2928$ eV on the superconductor only, calibrated to reproduce a zero-temperature gap of $0.17$~meV in an isolated 41-layer superconducting film, held fixed for all thicknesses and acting over the complete finite bands without an energy cutoff. 
Following \refnum{Tommaso2023}, we construct the linearized pairing kernel for zero in-plane pair momentum from the normal-state eigenstates of the full coupled stack, projected onto the superconducting orbitals. 
The two-dimensional Brillouin-zone average uses the measure $d^2k/(2\pi)^2$. 
The resulting $41\times41$ kernel permits a separate pairing amplitude on each superconducting layer; its largest eigenvalue $\lambda_{\max}>1$ signals a superconducting instability in this channel.
Momentum integration is refined until individual changes in the dimensionless kernel are below $10^{-3}$ in operator norm.
Within this model, $\lambda_{\max}$ reaches unity at approximately six spacer layers, marking the crossover from the normal to the superconducting regime (Extended Data Fig.~\ref{edf3}).

\textbf{Magnetization measurements.} 
Magnetization of the samples was measured using a Quantum Design MPMS3 SQUID magnetometer. The uniformity of the magnetic ordering across the epitaxial 2" wafer was verified by measuring a series of samples taken from the wafer edge to the center. After subtraction of the diamagnetic background, which was linear in magnetic field, the measured magnetic moment was close to the SQUID resolution of 10$^{-7}$ emu. For a sample volume of $3~\mathrm{mm}\times5~\mathrm{mm}\times12~\mathrm{nm}$, this corresponds to an  upper bound on the net magnetic moment of approximately $3\times10^{-3}~\mu_B$/Mn.
For comparison, a uniformly uncompensated interfacial Mn plane would correspond to a sample-averaged moment of approximately $0.1~\mu_B/\mathrm{Mn}$, using the $\sim0.3$~nm spacing of antiferromagnetically coupled Mn planes and the $\sim3.6~\mu_B$ Mn moment reported for tetragonal CuMnAs~\cite{Wadley2013, Wadley2017}.

\textbf{STEM characterization.}
The STEM lamellae were prepared using a focused-ion-beam scanning electron microscope (FIB-SEM) dual-beam system (Scios II). Electron-beam- and Ga-ion-beam-induced carbon deposition was used to protect the sample surface, and the final lamella was polished in the last step using a 2~keV Ga beam at 9~pA. HAADF-STEM combined with EDX was used for high-resolution imaging with simultaneous elemental mapping. These measurements were performed using a JEOL JEM NEOARM-200F microscope equipped with a Schottky-type field-emission gun operating at 200~keV. A JEOL JED-2300 EDX spectrometer was used for elemental analysis.

\textbf{XRD characterization.} 
X-ray diffraction measurements (Extended Data Fig.~\ref{edf1}) were performed using a Rigaku SmartLab diffractometer equipped with a 9~kW Cu rotating-anode source, a two-bounce Ge monochromator, and a HyPix detector.
An in-plane parallel-slit collimator limited the axial divergence, perpendicular to the vertical diffraction plane, to $1.0^\circ$. 
A $2.5^\circ$ Soller slit was placed in front of the detector to reduce contributions from air scattering. 
The measurements probed an illuminated sample area of several mm$^2$.
Radial scans were acquired using the detector in scanning 1D mode to reduce acquisition time. Reciprocal-space maps were recorded by combining multiple detector stripes measured at fixed scattering angle while varying the sample rocking angle. 
X-ray diffraction simulations were performed using xrayutilities within the kinematical multibeam framework, allowing substrate and multiple thin-film diffraction peaks to be described within a single model~\cite{kriegner2013xrayutilities}.

\textbf{Superconducting diode effect measurements.} 
Measurements were performed in a cryogen-free dilution refrigerator at a base temperature of 20 mK using a four-terminal configuration. 
A dc bias current was applied by sourcing a voltage from a Keithley 2614B SourceMeter across a 200~k$\Omega$ series resistor. 
The current through the device was measured using a current-to-voltage converter (Basel Precision Instruments, IF3602) and the four-terminal voltage using a voltage preamplifier (Basel Precision Instruments, SP1004). 
The amplified outputs were recorded with Agilent 34401A digital multimeters. 
A delay of 0.2~s was used between successive bias steps; this value was chosen empirically to minimize premature switching associated with faster current ramping. 
For the diode-efficiency measurements, the positive and negative switching currents were measured in separate sweeps and defined as the bias current at which the measured voltage first exceeded a threshold of 50~$\mu$V. 
For the zero-field-cooled switching-current distributions in Fig.~\ref{f3}a, 50 sweeps were acquired for each polarity, whereas for the other magnetic-field-evolution measurements the switching current for each polarity was averaged over eight sweeps.

\textbf{Nonlinear magnetotransport measurements.} 
Measurements were performed in a vector-magnet cryostat at 5~K. 
Devices for these measurements were fabricated from a CuMnAs wafer with 3 nm fully oxidized Al cap. 
The active device region was 20~$\mu$m long and 10~$\mu$m wide and was measured in a four-terminal configuration. 
The even-in-current nonlinear resistance was obtained from measurements at $+4$ mA and $-4$ mA as $R_{\mathrm{even}}=[V(+I)+V(-I)]/(2I)$. 
A Keithley 2400 SourceMeter was used to source the current, and a Keithley 2000 Multimeter was used to measure the longitudinal voltage. 

\textbf{XMLD-PEEM imaging.}
The X-ray magnetic linear dichroism photoemission electron microscopy measurements (Extended Data Fig.~\ref{edf5}) were performed at beamline I06 of Diamond Light Source. The X-ray beam was incident at a grazing angle of $16^\circ$, with the linear polarization lying in the plane of the film.
Spatially resolved XMLD contrast was obtained from the asymmetry, $\Delta=[I(E_2)-I(E_1)]/[I(E_2)+I(E_1)]$, between images acquired at the Mn $L_3$ absorption peak ($E_2$) and 0.9~eV below the $L_3$ peak ($E_1$), with a spatial resolution of approximately 30~nm.
The resulting contrast depends on the angle between the local spin axis and the X-ray linear-polarization direction. 
Images with the in-plane polarization axis oriented at $0^\circ$ and $45^\circ$ were obtained by rotating the sample with respect to the beam path. 
Measurements were performed at approximately 100~K using liquid-nitrogen cooling.
The polarization-dependent XMLD contrast indicates a preferential in-plane Néel-vector axis along [100], consistent with predominantly uniaxial magnetic anisotropy in the CuMnAs film.
\\

\textbf{ACKNOWLEDGEMENTS}

Research was supported by research grants from VILLUM FONDEN (Project Nos. 43951 and 53097), Lumina Quaeruntur fellowships LQ100102201 and LQ100102602 of the Czech Academy of Sciences, the Czech Science Foundation (Grant Nos. 22-22000M and 25–18281K), the Ministry of Education of the Czech Republic Grant No. CZ.02.01.01/00/22008/0004594 and e-INFRA CZ (ID:90254), and co-funded by the European Union Physics for Future (Grant No. 101081515). 
We acknowledge CzechNanoLab Research Infrastructure supported by MEYS CR (LM2023051) and OP VVV "Excellent Research Teams", project No. CZ.02.1.01/0.0/0.0/15$\_$003/0000417 – CUCAM, as well as ERC Starting Grant No. 101165122, and ERC Advanced Grant No. 101095925.\\

\textbf{CONTRIBUTIONS} 

FK and SV conceived the project. FK, VN and JK developed and performed the material growth. FK, AF and MM performed the STEM characterization. JK and DK performed the XRD characterization. OA, KE and PW performed the XMLD-PEEM measurements. TE, AJ, AS, SV, CM and LC performed the device fabrication, superconducting transport characterization and SDE measurements. LŠ, ABH, VP, SG and NH performed the DFT and superconducting-instability calculation. KO performed the device fabrication and nonlinear magnetotransport measurements. FK, TJ and SV wrote the manuscript with input from all co-authors. \\

\newpage
\textbf{EXTENDED DATA}\\

\setcounter{figure}{0}
\renewcommand{\figurename}{Extended Data Fig.}

\renewcommand{\figurename}{Extended Data Fig.}
\renewcommand{\thefigure}{\arabic{figure}}
\renewcommand{\theHfigure}{ED.\arabic{figure}}

%%%%%%%%%%%%%%%%%%%%% ED FIG. 1 %%%%%%%%%%%%%%%%%%%%%%%
\begin{figure*}[h!]
    \centering
    \includegraphics[width = 1\linewidth]{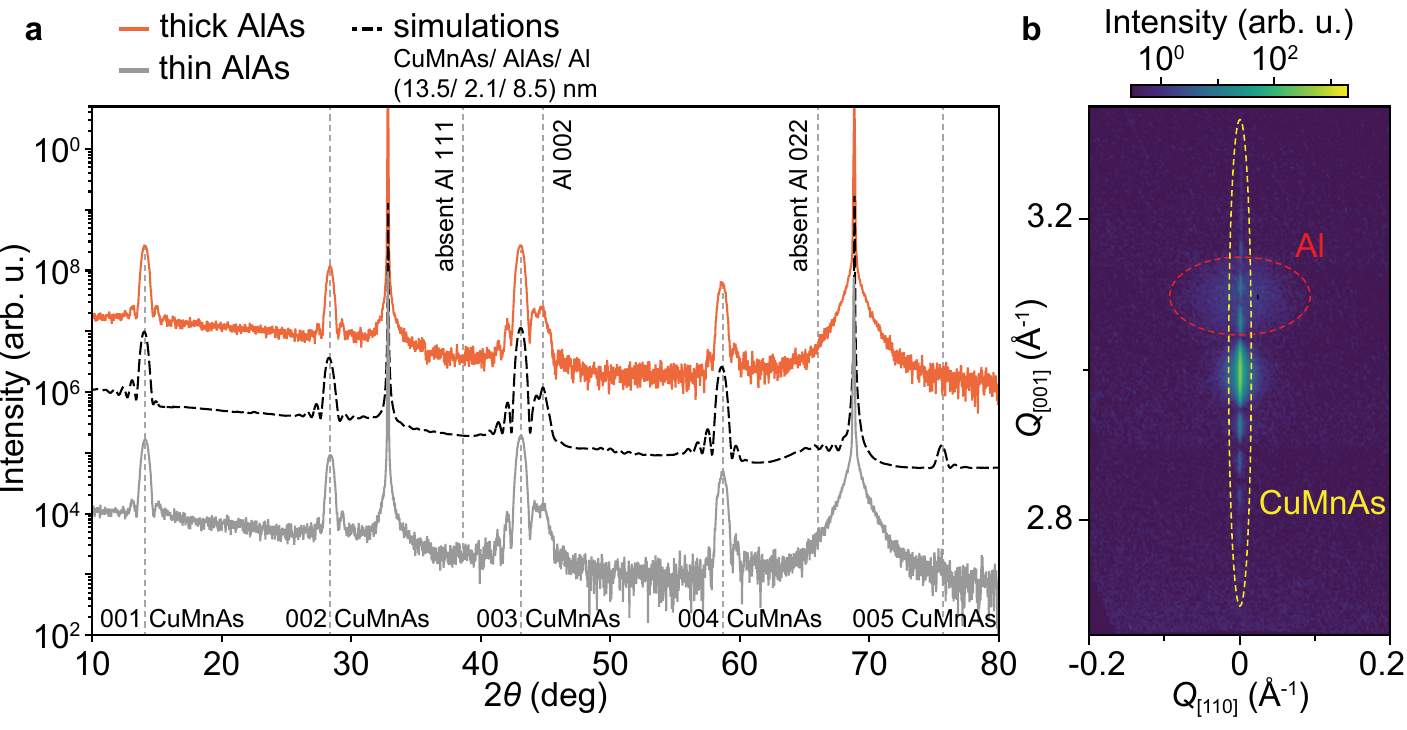}
    \caption{ 
    \textbf{XRD characterization.}
    \textbf{a,} Radial scan of the CuMnAs heterostructure with the thicker AlAs barrier (orange) and the reference sample with the thinner AlAs barrier (gray). 
    The black dashed line shows a kinematical multibeam simulation~\cite{kriegner2013xrayutilities} for the indicated CuMnAs/AlAs/Al layer thicknesses, including some incoherent averaging for small thickness variations. 
    The close agreement with the experimental diffraction patterns confirms that Al epitaxially matches both CuMnAs and AlAs, with Al 002 aligned along the heterostructure growth direction. 
    The absence of the theoretically stronger Al 111 and 022 peaks indicates a well-defined single 001 orientation of the Al film. 
    \textbf{b,} Reciprocal space map around the CuMnAs 003 and Al 002 peaks, indicating strong in-plane lattice alignment of the CuMnAs, AlAs, and Al layers.
    }
    \label{edf1}
\end{figure*}
%%%%%%%%%%%%%%%%%%%%%%%%%%%%%%%%%%%%%%%%%%%%%%%%%%%%

%%%%%%%%%%%%%%%%%%%%% ED FIG. 2 %%%%%%%%%%%%%%%%%%%%%%%
\begin{figure*}[h!]
    \centering
    \includegraphics[width = 1\linewidth]{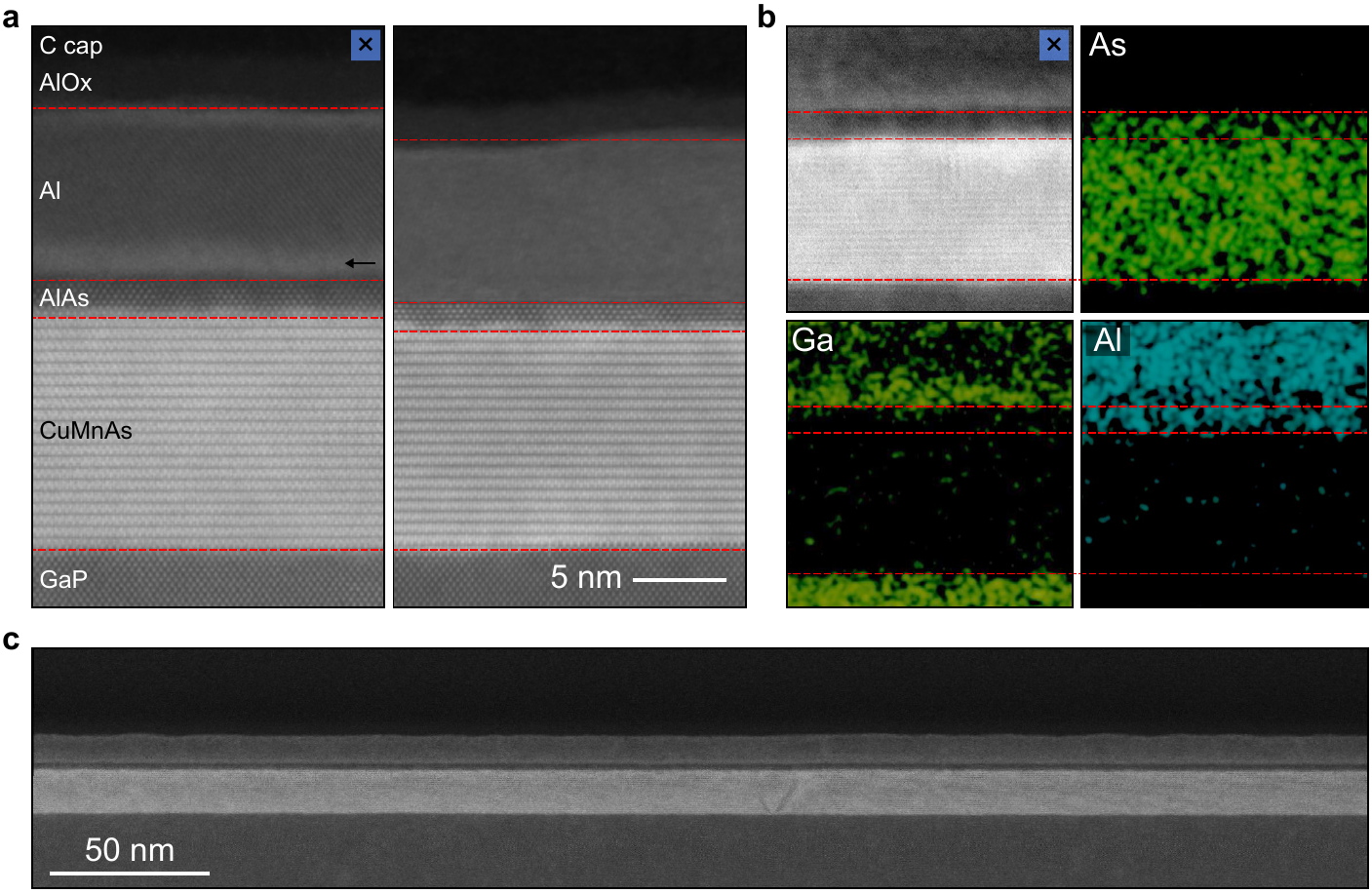}
    \caption{ 
    \textbf{STEM and EDX characterization.} 
    \textbf{a,} HAADF-STEM images of the CuMnAs/AlAs/Al heterostructure and the reference sample with the thinner AlAs barrier. 
    The measured layer thicknesses in the samples are 9.2~nm Al, 2.0~nm AlAs, and 12.4~nm CuMnAs for the main heterostructure, and 8.7~nm Al, 1.55~nm AlAs, and 11.7~nm CuMnAs for the reference sample. 
    The thicknesses were extracted using the red dashed guides, which were placed at the extrema of the integrated, interpolated vertical intensity gradient. 
    \textbf{b,} Energy-dispersive X-ray spectroscopy of the main heterostructure. 
    The spatial distributions of As, Ga, and Al confirm the chemical composition of the AlAs barrier. 
    Additionally, we detect the presence of Ga atoms in the Al layer, with an enhanced concentration at the AlAs/Al interface. 
    We attribute this to damage introduced by Ga-based focused-ion-beam lamella preparation. 
    The as-grown single-crystalline nature and high degree of epitaxy of Al to the AlAs barrier are supported by the XRD data in Extended Data Fig.~\ref{edf1}. 
    \textbf{c,} A HAADF-STEM overview of a larger region of the main heterostructure, highlighting the homogeneity and quality of the layer stack. 
    }
    \label{edf2}
\end{figure*}
%%%%%%%%%%%%%%%%%%%%%%%%%%%%%%%%%%%%%%%%%%%%%%%%%%%%

%%%%%%%%%%%%%%%%%%%%% ED FIG. 3 %%%%%%%%%%%%%%%%%%%%%%%
\begin{figure*}[h!]
    \centering
    \includegraphics[width = 0.8\linewidth]{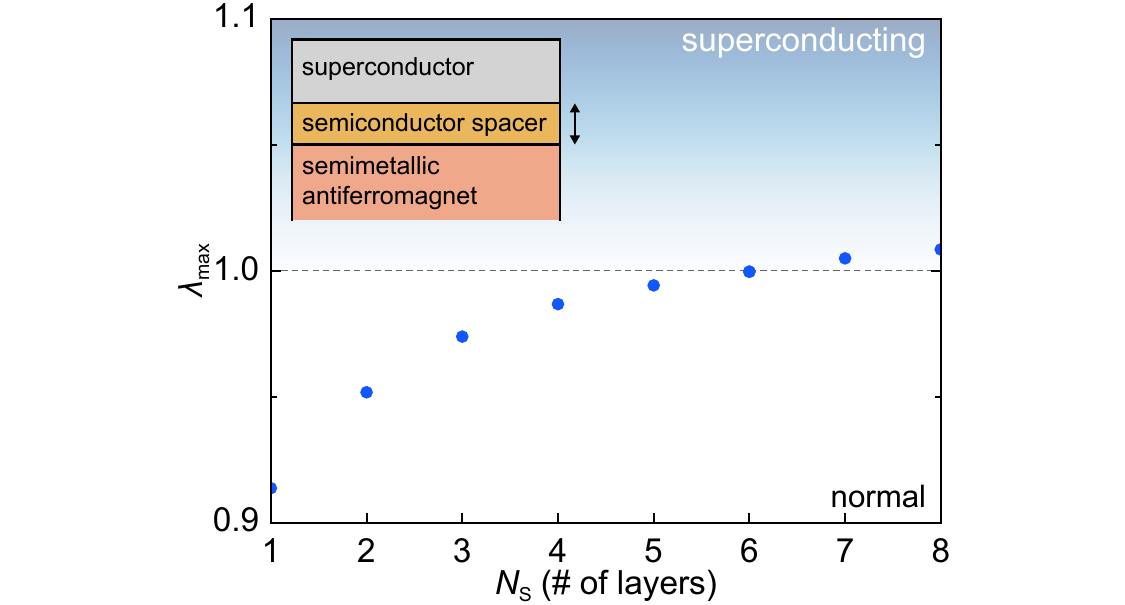}
    \caption{\textbf{Superconducting instability versus spacer thickness.}
    Largest eigenvalue $\lambda_{\max}$ of the linearized pairing kernel as a function of the number $N_{\rm S}$ of semiconductor spacer layers at $T=1$~K. 
    The superconducting region contains 41 layers, while the antiferromagnet is represented by a single effective layer. 
    Values $\lambda_{\max}>1$ indicate an instability of the normal state toward onsite singlet superconductivity. 
    All other model parameters are held fixed.
    The instability threshold is reached at approximately six spacer layers. Taking one (001) AlAs monolayer as approximately $0.283~\mathrm{nm}$, the experimentally studied $1.5$ and $2.0~\mathrm{nm}$ barriers correspond to approximately five and seven spacer layers, respectively, placing them on opposite sides of the calculated crossover. See Methods for model and calculation details.}
    \label{edf3}
\end{figure*}
%%%%%%%%%%%%%%%%%%%%%%%%%%%%%%%%%%%%%%%%%%%%%%%%%%%%

%%%%%%%%%%%%%%%%%%%%% ED FIG. 4 %%%%%%%%%%%%%%%%%%%%%%%
\begin{figure*}[h!]
    \centering
    \includegraphics[width = 0.8\linewidth]{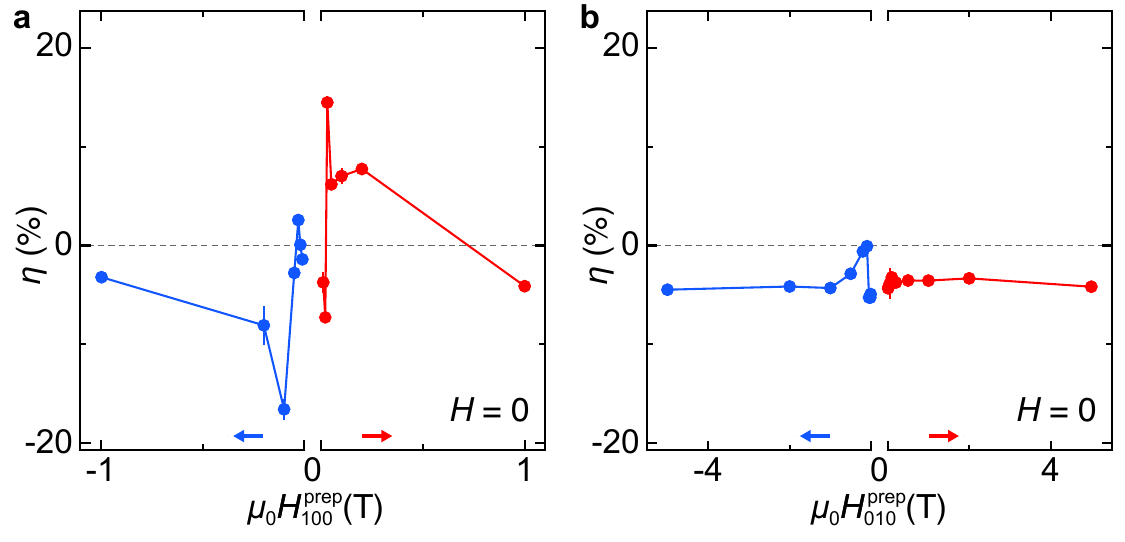}
    \caption{ 
    \textbf{Superconducting diode response of the [100]-oriented bar.} \textbf{a, b,} Zero-field superconducting diode efficiency, $\eta$, measured after magnetic-field preparation along \textbf{a,} $[100]$ and \textbf{b,} $[010]$ crystallographic directions. 
    The device is measured at $H=0$ after each preparation field, following the same protocol as in Fig.~\ref{f3}c. 
    The response shows the same qualitative dependence on the preparation-field axes as the $[010]$-oriented bar, suggesting that the anisotropy is not determined simply by the relative orientation of the applied field and current, but is instead tied to the underlying crystallographic axes of CuMnAs.
    }
    \label{edf4}
\end{figure*}
%%%%%%%%%%%%%%%%%%%%%%%%%%%%%%%%%%%%%%%%%%%%%%%%%%%%

%%%%%%%%%%%%%%%%%%%%% ED FIG. 5 %%%%%%%%%%%%%%%%%%%%%%%
\begin{figure*}[h!]
    \centering
    \includegraphics[width = 0.8\linewidth]{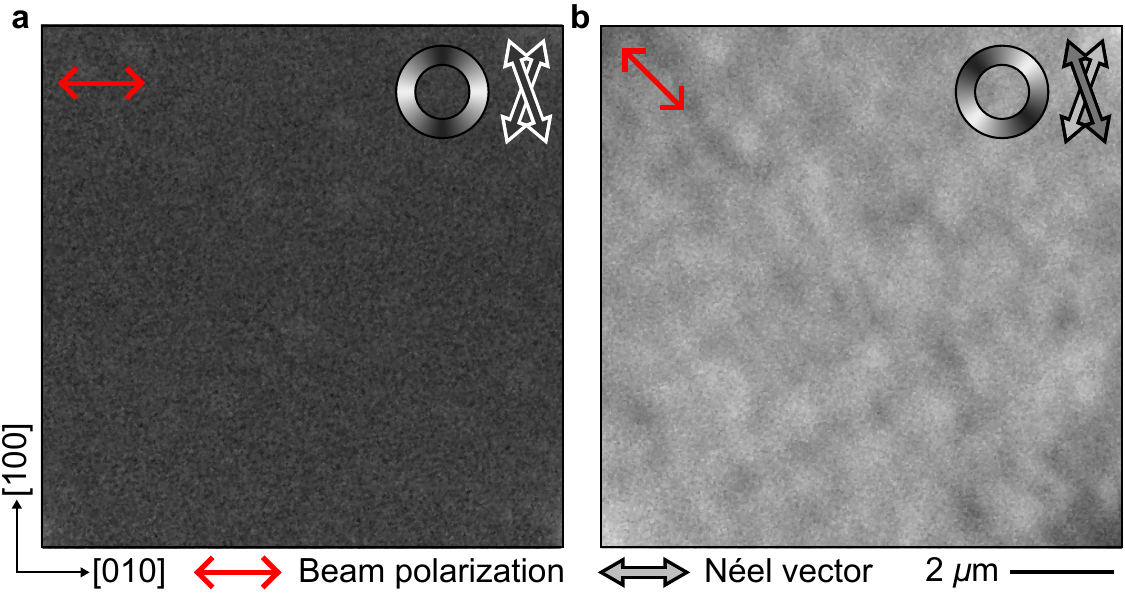}
    \caption{ 
    \textbf{XMLD-PEEM imaging of the CuMnAs Néel texture.} 
    \textbf{a, b} XMLD-PEEM images of a $10~\mu\mathrm{m}\times10~\mu\mathrm{m}$ region of the CuMnAs epilayer. 
    Red arrows indicate the in-plane X-ray linear-polarization direction, and double-headed arrows indicate the Néel-vector axis.
    \textbf{a,} X-ray polarization along $[010]$. 
    \textbf{b,} X-ray polarization rotated by $45^\circ$ from $[010]$. 
    The XMLD intensity depends on the local orientation of the Néel-vector axis (greyscale arrows) relative to the X-ray polarization (red arrows), as indicated by the colour wheels.
    }
    \label{edf5}
\end{figure*}
%%%%%%%%%%%%%%%%%%%%%%%%%%%%%%%%%%%%%%%%%%%%%%%%%%%%

\clearpage

\bibliography{papers_1}

\end{document}